\documentclass[aps,prc,twocolumn,groupedaddress,showpacs]{revtex4}

\usepackage{graphicx}
\usepackage{dcolumn}
\usepackage{bm}
\usepackage{amssymb,amsmath}

\def\F{{\cal F}}
\def\GV{G_{\mbox{\tiny V}}}

\def\DRV{\Delta_{\mbox{\tiny R}}^{\mbox{\tiny V}}}

\begin{document}

\preprint{Cyclotron Institute at Texas A\&M University}

\title{Precise Half-Life Measurement of the \\
	Superallowed $\beta^+$ Emitter $^{26}$Si}

\author{V.E. Iacob}
\email[]{iacob@comp.tamu.edu}
\altaffiliation {On leave from the National Institute for Physics and Nuclear Engineering ``Horia Hulubei", Bucharest, Romania.}
\author{J.C. Hardy}
\author{A. Banu}
\altaffiliation {On leave from the National Institute for Physics and Nuclear Engineering ``Horia Hulubei", Bucharest, Romania.}
\author{L. Chen}
\author{V. V. Golovko}
\altaffiliation {Present address: Department of Physics, Queen's University, Stirling Hall, Kingston, ON, K7L 3N6, Canada.}
\author{J. Goodwin}
\author{V. Horvat}
\author{N. Nica}
\altaffiliation {On leave from the National Institute for Physics and Nuclear Engineering ``Horia Hulubei", Bucharest, Romania.}
\author{H.I. Park}
\author{L. Trache}
\author{R.E. Tribble}
\affiliation{Cyclotron Institute at Texas A\&M University}

\date{\today}

\begin{abstract}

We have measured the half-life of the superallowed $0^+$$\rightarrow$\,$0^+$ $\beta^+$\,emitter $^{26}$Si
to be 2245.3(7)\,ms.  We used pure sources of $^{26}$Si and employed a high-efficiency gas counter, which
was sensitive to positrons from both this nuclide and its daughter $^{26}$Al$^m$.  The data were analyzed
as a linked parent-daughter decay.  To contribute meaningfully to any test of the unitarity of the
Cabibbo-Kobayashi-Maskawa (CKM) matrix, the $ft$ value of a superallowed transition must be determined to a
precision of 0.1\% or better.  With a precision of 0.03\% the present result is more than sufficient to be
compatable with that requirement.   Only the branching ratio now remains to be measured precisely before a
$\pm$0.1\% $ft$ value can be obtained for the superallowed transition from $^{26}$Si.

\end{abstract}

\pacs{21.10.Tg,23.40.-s,27.30.+t}

\maketitle

\section{Introduction \label{intro}}

The unitarity of the Cabibbo-Kobayashi-Maskawa (CKM) matrix is a fundamental
requirement of the three-generation Standard Model.  Currently, the most demanding
test available of CKM unitarity is the sum of squares of the experimentally
determined elements of the matrix's top row \cite{Ha09}. The dominant term
in this test is the up-down quark-mixing element, $V_{ud}$, the most precise value
of which is obtained through nuclear measurements of superallowed $0^+$$\rightarrow$\,$0^+$
beta decays. To date, the measured $ft$ values for transitions from ten different nuclei are
known to $\sim$0.1\% precision and three more to $\leq$0.3\% \cite{Ha09}.  So far, the
superallowed transition from $^{26}$Si has not been in either category, its $ft$-value
precision being 0.8\%, too large for it to contribute to the unitarity test.

Since a superallowed $0^+$$\rightarrow$\,$0^+$ transition involves only the vector current, its
$ft$ value relates to the vector coupling constant, $\GV$ -- and, through it, to $V_{ud}$ --
via the relationship \cite{Ha09}
\begin{equation}
\F t \equiv ft (1 + \delta_R^{\prime}) (1 + \delta_{NS} - \delta_C ) = \frac{K}{2 \GV^2 
(1 + \DRV )}~,
\label{Ftconst}
\end{equation}
where $\F t$ is defined to be the ``corrected" $ft$ value and $K/(\hbar c )^6 = 2 \pi^3 \hbar \ln 2
/ (m_e c^2)^5 = 8120.2787(11) \times 10^{-10}$ GeV$^{-4}$s.  There are four small correction terms: $\delta_C$
is the isospin-symmetry-breaking correction; $\DRV$ is the transition-independent part of the radiative
correction; and the terms $\delta_R^{\prime}$ and $\delta_{NS}$ comprise the transition-dependent part
of the radiative correction, the former being a function only of the maximum positron energy and the atomic number, $Z$, of
the daughter nucleus, while the latter, like $\delta_C$, depends in its evaluation on the details of
nuclear structure.  Both $\delta_C$ and $\delta_{NS}$ have been calculated \cite{To08} with the best
available shell-model wave functions, which are based on a wide range of spectroscopic data.  They include
those core orbitals that were determined to be important based on measured spectroscopic factors in
single-nucleon pick-up reactions; and they were further tuned to agree with measured binding energies, 
charge radii and coefficients of the isobaric multiplet mass equation.

Although, in principle, the precise $ft$ value for a single $0^+$$\rightarrow$\,$0^+$ superallowed transition
should be sufficient to determine $V_{ud}$, that would leave the validity of these structure-dependent
corrections without independent verification, and the derived value for $\GV$ could be in doubt.  What
has given credibility to the nuclear result for $\GV$ is the fact that many superallowed transitions have
been measured precisely and all give statistically identical results for $\F t$ -- and hence for $\GV$.  
Since the uncorrected $ft$ values actually scatter over a relatively wide range, it is the structure-dependent
corrections that are responsible for bringing the $\F t$ values into agreement with one another.

Obviously, this is already a powerful experimental validation of the calculated corrections themselves, but
it can be improved even further by precise measurements of additional transitions, especially ones calculated to have
large correction terms.  If the $ft$ values measured for cases with large calculated corrections also turn
into corrected $\F t$ values that are consistent with the others, then this reinforces the calculations'
reliability for cases in which the corrections are smaller.  The calculated correction terms for the superallowed
transition from $^{26}$Si give the result $\delta_C-\delta_{NS}$ = 0.65(3)\% \cite{To08}.  Although this value
is not particularly large compared to most of the well-measured cases, it is more than double the size of
the correction for the superallowed transition from $^{26}$Al$^m$, which is its mirror transition.
Experimental verification of this predicted mirror asymmetry would be a valuable test of the calculations.

The $ft$ value that characterizes any $\beta$-transition depends on three measured quantities: the
total transition energy, $Q_{EC}$, the half-life, $t_{1/2}$, of the parent state and the branching ratio,
$R$, for the particular transition of interest.  The $Q_{EC}$-value is required to determine the statistical
rate function, $f$, while the half-life and branching ratio combine to yield the partial half-life, $t$.
The $Q_{EC}$ value for $^{26}$Si is already known sufficiently well \cite{Er09} to yield a value for $f$ with
0.015\% precision but, before the measurement reported here, the $^{26}$Si half life was only known to 0.12\%
and its branching ratio to 0.8\% \cite{Ha09}.  Our new half-life for $^{26}$Si has 0.03\% precision and furthermore
disagrees significantly with the measurement \cite{Ma08} that previously dominated the world average.  The
result reported here represents our first step in bringing the precision of the $ft$ value for this transition into
the desired range of 0.1\%.  The second step will be an improved branching ratio, a measurement that we will soon undertake.

\begin{figure}
 \includegraphics[width=\columnwidth]{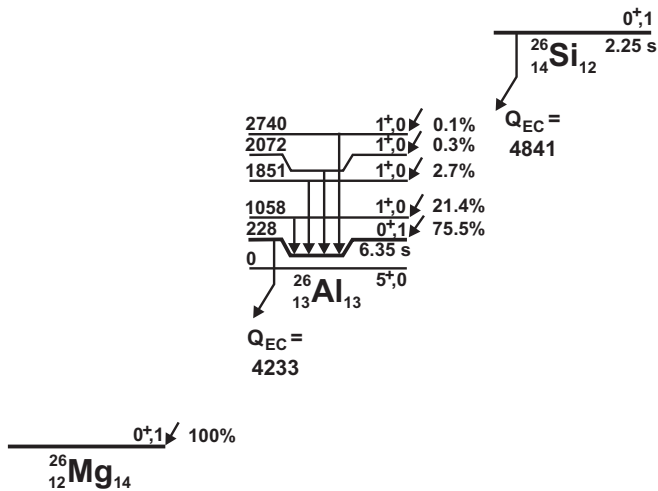}%
 \caption{\label{fig1} Combined decay schemes of $^{26}$Si and $^{26}$Al$^m$ showing only those features of relevance
to their superallowed $\beta$ decays.  All energies are in keV and the $Q_{EC}$ values shown are for the superallowed
branch.  The data are taken from Refs.\,\cite{En90,Ha09}.}
\end{figure}

Like the decay of $^{34}$Ar, whose half-life we have reported previously \cite{Ia06}, $^{26}$Si $\beta^+$ decays
to a daughter which is itself radioactive and is, in fact, another superallowed emitter.  The combined decay schemes
appear in Fig.\,\ref{fig1}, where it can be seen that the half-life of $^{26}$Al$^m$, the daughter, differs by only
a factor of three from that of $^{26}$Si.  This adds complications to the experiment, which requires us to use the
techniques that we developed previously for our study of $^{34}$Ar.  These will be described briefly in the following
sections, but for further details the reader is referred to Ref.\,\cite{Ia06}.

\section{Experiment \label{exp}}
%
\subsection{Overview \label{overview}}
%
Precise half-life measurements need high-purity radioactive beams, a requirement that is even more important
for cases where the daughter nucleus is also radioactive with a similar half-life. We achieved this goal by
using a production reaction with inverse kinematics, $p$($^{27}$Al,\,$2n$)$^{26}$Si, and selecting the desired
reaction product with the Momentum Achromat Recoil Separator (MARS) \cite{MARS2002}. Our primary beam of
30$A$-MeV $^{27}$Al, which was produced by the superconducting cyclotron at Texas A\&M University, impinged on
a 1.6-atm hydrogen gas target cooled to liquid nitrogen temperature.  The fully stripped ejectiles were then
analyzed by MARS.  Initially, working with a low-current primary beam, we inserted at the focal plane of MARS
a 5$\times$5 cm silicon telescope consisting of a 16-strip position-sensitive detector (PSD) 300 $\mu $m
thick, backed by a 1-mm-thick detector. The telescope was used first for the identification of secondary
reaction products, then for the control of the selection and focus of the desired species in the center
of the beam line.  This also gave us a clear indication of nearby reaction products that could potentially
contribute as impurities to our selected beam.
 
After the tuning and selection procedure, the PSD was dropped out of the way and the intensity
of the primary beam was increased.  With extraction slits at the MARS focal plane used to select $^{26}$Si,
the resulting 25.2-$A$-MeV radioactive beam was extracted into air through a 51-$\mu$m-thick kapton window.  
This beam, typically of $2\times10^4$ ions per second, passed through a $0.3$-mm thin BC-404 plastic scintillator, 
where the ions were counted, and then through a set of aluminum degraders, eventually being implanted in the
76-$\mu$m-thick aluminized mylar tape of a fast tape-transport system.  The combination of $m/q$ selectivity
in MARS and range selectivity in the degraders led to implanted samples that were at least 99.8\% pure.

After the radioactive sample had been collected for a time interval of the order of the $^{26}$Si half-life, the beam was
turned off and the tape-transport system moved the sample in $\sim$175\,ms to a well-shielded location 90\,cm away, 
stopping it in the center of a 4$\pi$ proportional gas counter.
The decay positrons were then detected for twenty half-lives (45 s), with signals from the gas counter
being multiscaled into a 500-channel time spectrum. These collect-move-detect cycles were computer controlled and
their timing was continuously monitored on-line.  They were repeated, with a separate decay spectrum recorded for each, 
until the desired overall statistics had been achieved.  In its shielded location, the gas counter had a background
rate of about 0.5 counts/s, which was 3-4 orders of magnitude lower than the initial count rate for each
collected sample.  For this experiment we accumulated data from 5,000 cycles split into 55 separate runs, which
yielded a total of $2 \times 10^8$ counts.

\subsection{Gas counter and electronics \label{gasc}}
%
The gas counter we used is similar to ones we have used in previous precise half-life measurements \cite{Ko83,Ko97,Ia06,Ha03,Ia08}.
It consists of two separate gas cells that, when assembled, have a 0.25-mm slot between them, through which the
mylar transport-tape passes.  Both cells were machined from copper and each is equipped with anodes of 13-$\mu$m-diameter
gold-plated tungsten wire.  Methane at just over one-atmosphere pressure is continuously flushed through both cells.  Methane
offers adequate gas gain for detecting positrons and is quite insensitive to $\gamma$ radiation.  A Havar foil window, 3.7 cm
in diameter and 1.5 $\mu$m thick, hermetically seals each gas cell on the side facing the tape.

The electronic chain we used in the measurement was the same as that described in detail in Ref.\,\cite{Ia06}.  The
pre-amplified signal from the gas counter is first passed to a fast-filter amplifier with high gain ($\times$500).  At this
high gain many of the pulses would saturate the amplifier so, to ensure that the amplifier recovers quickly, large pulses are
clipped with a Schottky diode inserted after its first stage of amplification.  The amplified and clipped pulses are then
passed to a discriminator with very low threshold (150-250 mV).

A $^{90}$Sr/$^{90}$Y $\beta$ source is used to set up the detector system.  This source has been specially prepared
on a sample length of transport tape and is inserted into the gas counter in exactly the position that an on-line
sample, such as $^{26}$Si, is situated.  Our procedure is to record, at a fixed threshold setting, the counting
rate from the discriminator as a function of the applied counter bias voltage.  Initially, as the applied voltage
is raised the count-rate also rises since the increasing gas gain leads to more
primary ionizing $\beta$ events triggering the discriminator.  However, at approximately 2600 volts -- the exact value
depends on the threshold setting -- a ``plateau" is reached, and the count rate remains nearly unchanged for the next
200-300 volts increase in the bias voltage.  At higher voltages still, there is a second rapid rise in the count rate
as spurious pulses increasingly trigger the discriminator.  This behavior is well understood \cite{NC85} and clearly
demonstrates that, when operated in the plateau region, such detectors have essentially 100\% efficiency.  During our
$^{26}$Si measurement the detector was always operated in the plateau region.   

Since dead-time is a serious concern for half-life measurements, the discriminator output signals were split and sent
to two fixed-width, non-retriggering and non-extending gate generators, which established different dominant dead
times in the two separate streams, both of which were multiscaled.  The time base for the multiscalers was defined
by a function generator, which is accurate to 0.01 ppm.  Both gates also were continuously monitored during every run, 
thus giving us an on-line measure of the dead-time ($\pm$5 ns) during data collection.  Note that even though the two
gate generators were fed by the same data, the dead-time distortions of the underlying Poisson-distributed data are
independent in the two cases; thus the two data streams allowed us to test that our dead-time corrected result was
independent of the actual dead time of the circuit.

\subsection{Special precautions \label{prec}}
%
As the experiment was aimed at better than 0.1\% precision, many tests for systematic effects were made and special
precautions taken during the measurements themselves:

\begin{itemize}

\item 	Every experiment was subdivided into many separate runs, differing only in their particular combination of
detection parameters: dominant dead-time, detector bias and discrimination threshold.  We used combinations of four
different dead times (3, 4, 6 and 8 $\mu$s), three discriminator thresholds (150, 200 and 250 mV) and four detector
biases (2550, 2650, 2750 and 2850 V).  The separate analysis of each individual run allowed us to test for systematic
effects that could contribute to the uncertainty in the final result.

\item   Since each $^{26}$Si decay produces an $^{26}$Al$^m$ daughter that also decays, the relative activity of the two
nuclides at the beginning of the detection period depends on the length of the collection period (and the tape-move
time) that preceded it.  We used four different collection times (1.0, 2.5, 3.0 and 4.5 s) to test for consistency.

\item   The ratio of the parent to daughter activities also depends on the time-dependence of the rate
at which $^{26}$Si was accumulated in the tape during the collection period.  The number of ions registered in the
scintillator located just in front of the aluminum degraders was recorded as a function of time with each cycle, and
the results were used in our analysis.

\item   The few weak impurities in the $^{26}$Si beam have different ranges in our degraders.  Thus, any contribution
they might make to the total activity collected in the tape would depend on the depth at which the $^{26}$Si beam
is stopped in the tape.  By using two different thicknesses of aluminum degrader we placed the $^{26}$Si midway
through the 76-$\mu$m tape for the first 30 runs and then near the rear surface for the remaining 25.  Again we tested for
consistent results.

\item   The tape-transport system is quite consistent in placing the collected source within
$\pm$3 mm of the center of the detector, but it is a mechanical device, and occasionally larger
deviations occur.  We separately recorded the number of implanted nuclei detected in the scintillator
during the collection period of each cycle, and the number of positrons recorded in the gas counter during the
subsequent count period.  The ratio of the latter to the former is a sensitive measure of whether
the source was seriously misplaced in the proportional counter.  

\item   We checked the composition of the beam exiting MARS once per day by re-inserting the PSD and ensuring that
no changes had occurred.

\item   Several background measurements were made, in which all conditions were identical to those of a normal run
except that the tape motion was disabled.

\item   In one run, we repeatedly collected activity for 16.5 s and counted for 165 s in order to search for
long-lived impurities.  None was found.

\item   It is important that the gas counter be operated in the ``plateau" region: {\it i.e.}~within the range of
detector bias voltages in which the counting rate is nearly independent of voltage at a given discriminator
setting.  We measured this voltage plateau with a long-lived $^{90}$Sr/$^{90}$Y source before and after our
measurement, and once during it.  In all cases we found the plateau slope to be $\leq$0.5\% per 100 V and did
not observe any changes in the voltage boundaries of the plateau region.
\end{itemize}

\section{Analysis and Results \label{res}}
Before analyzing the data, we first removed any cycles that had fewer than 500 implanted $^{26}$Si ions detected
by the scintillator.  These were the result of low -- or no -- primary beam current from the cyclotron during part
or all of the collection period.  Then we eliminated those cycles with an anomalously low ratio of recorded positrons
to implanted silicon ions, which is indicative of faulty tape motion leading to a misplaced sample in the gas detector.  
Approximately 8\% of the cycles were rejected for this reason.

\subsection{Parent-daughter connection \label{sub:pd}}

\begin{figure}[t]
  \includegraphics[width=\columnwidth]{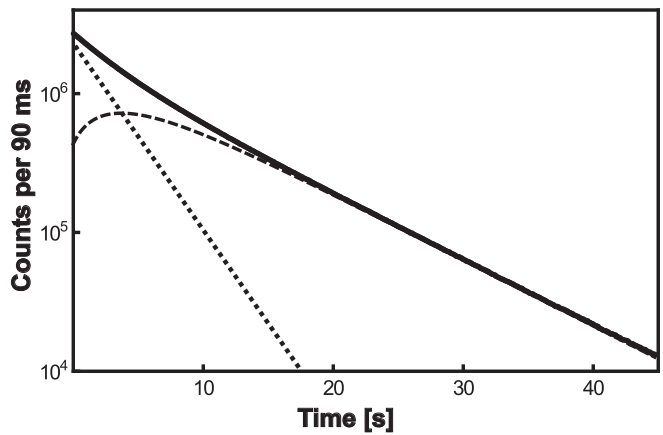}%
 \caption{\label{fig2} Measured time-decay spectrum (solid line) for the total of all data obtained from the
 $\beta^+$ decay of $^{26}$Si and its daughter $^{26}$Al$^m$.  The dotted/dashed lines represent the derived
 $^{26}$Si/$^{26}$Al$^m$ contributions.}
\end{figure}

For each run, we processed the accepted data by summing the dead-time-corrected spectra from all its
included cycles.  We corrected for dead time using the procedure outlined in Ref.\,\cite{Ko97}.  The
total time-decay spectrum obtained from the combined runs is presented in Fig.\,\ref{fig2}, 
where we also show the separate contributions from the $^{26}$Si parent and $^{26}$Al$^m$
daughter.  This breakdown into components is based upon our final analysis and
is presented here simply to illustrate how the parent-daughter decay curve, which combines two rather similar
half-lives, tends to mask the parent half-life even though the parent activity dominates at the start of
the counting period.

We can easily understand this situation by examining the coupled decay equations for combined parent-daughter
decays. The combined $^{26}$Si and $^{26}$Al$^m$ activity yields a total rate for detected positrons of 
\begin{equation}
\Lambda _{tot}  =
	C_1 \, e^{ - \lambda _1 t}  + 	C_2 \, e^{ - \lambda _2 t} , \label{2comp}
\end{equation}
with
\begin{eqnarray}
 C_1  &=& N_1 \epsilon_2 \lambda _1 \left (\frac{\epsilon_1}{\epsilon_2} - \frac{\lambda _2}{\lambda _1  - \lambda _2 }\right ) \nonumber \\ 
 C_2  &=& N_1 \epsilon_2 \lambda_2 \left ( \frac{N_2}{N_1} + \frac{\lambda _1 }{\lambda _1  - \lambda _2 } \right ) , \label{coeffs}
 \end{eqnarray}

\noindent where $t$ is the time elapsed after the end of the collect period; $N_{1,2}$ are the numbers
of $^{26}$Si and $^{26}$Al$^m$ nuclei present in the sample at $t = 0$; $\epsilon_{1,2}$ are the experimental
efficiencies for detecting the positrons from the respective decays; and $\lambda _{1,2}$ are the corresponding
decay constants. Note that when $\lambda _1 = 2\lambda _2$ (and $\epsilon_1 = \epsilon_2$) the coefficient $C_1$
vanishes, leaving a single exponential term having the decay constant of the daughter. The half-lives of $^{26}$Si
and $^{26}$Al$^m$ are actually related by a factor of 2.8, close enough to 2 that, for our measurements, the
coefficient $C_1$ was slightly smaller than $C_2$, leaving the daughter to dominate the decay curve (see
Fig.\,\ref{fig2}).  This imposes a real limitation on the precision that can be obtained from a conventional
fit to the data: even with $\lambda_2$ fixed at its known value, $C_1$, $C_2$ and $\lambda_1$ (as well as
the constant background) must all be determined independently, which leads to $\lambda_1$ having a rather
large uncertainty.

We can overcome this limitation by fixing the ratio $C_2/C_1$ so as to reduce the number of adjustable parameters
in the fit to three (including background), as we did for our measurement of the $^{34}$Ar half-life \cite{Ia06}.  
In practice this means that we must establish two ratios, $N_2/N_1$ and $\epsilon_1/\epsilon_2$, from the
experimental parameters.  In this section we deal with the former; in the next section, with the latter.

\begin{figure}[t]
  \includegraphics[width=\columnwidth]{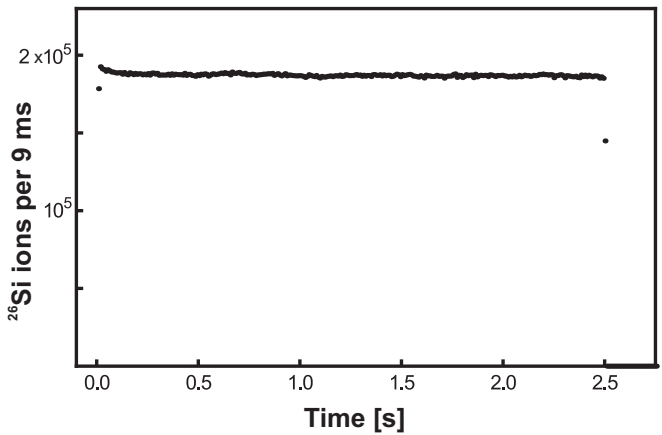}
  \caption{\label{fig3} Typical time-profile of the collected $^{26}$Si beam. The initial drop in
intensity is generated by the change in gas density as the primary beam heats the gas around its path.  
A fan located inside the gas-target ensures a rapid transition to stable conditions. 
	}
\end{figure}

No $^{26}$Al ions were present in the $^{26}$Si sample collected in our tape (see Sec.\,\ref{sub:imp}).  Thus, if the sample
collection rate were exactly constant, we could determine $N_2/N_1$ from a simple calculation of the production of
$^{26}$Al$^m$ (via $^{26}$Si decay) over the precisely known time of the collection period.  However, we measured
the actual number of $^{26}$Si ions as a function of time with the scintillator at the exit of MARS and determined
that there was a slightly higher rate at the very beginning of the collection period (see
Fig.\,\ref{fig3}).  Our cryo-target was a gas, and the primary beam heats that gas around its path through the target, 
thus initially generating a local drop in gas density; the transition to steady conditions was hastened by a fan
that continuously circulated the gas in the target cell.  We also found that the size of the variation in the
radioactive-beam intensity at the beginning of a cycle depended on the primary beam intensity, potentially changing
the beam-profile from one cycle to another.  Nevertheless, with the collection time-profile measured and recorded
for each cycle, we could perform a numerical integration over the measured $^{26}$Si accumulation to calculate the
decay-production of $^{26}$Al$^m$ and the corresponding $N_2/N_1$ ratio accurately for each run.

\subsection{Parent-daughter relative efficiencies \label{sub:eff}}

\begin{figure}
  \includegraphics[width=7.5cm]{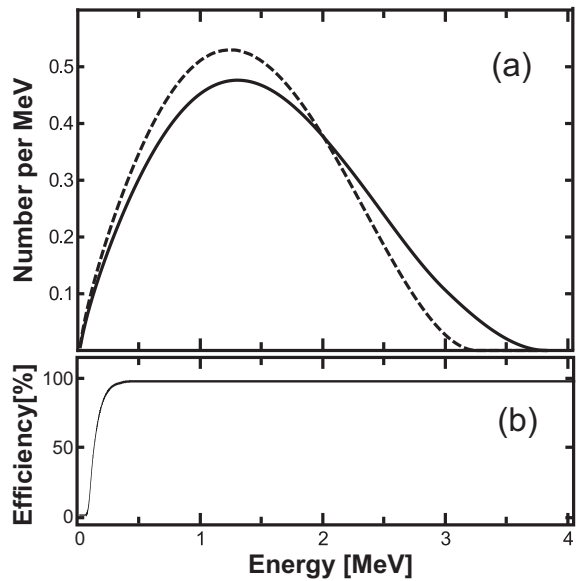}
 \caption{\label{fig4} (a) Calculated $\beta^+$ spectra for $^{26}$Si (solid curve) and $^{26}$Al$^m$ (dashed curve).  The
former includes the Gamow-Teller branches to $1^+$ states as well as the superallowed branch to the $0^+$ isomer; the latter
is pure superallowed decay (see Fig.\,\ref{fig1} for both decay schemes).  (b) System efficiency for detecting positrons due
to the effects of the aluminized Mylar tape and the Havar windows of the detector gas cells.  The curve is the result of a
Monte Carlo calculation.}
\end{figure}

The ratio $\epsilon_1/\epsilon_2$ can also be established
independently.  In our experiment, we detected positrons from the decays of both $^{26}$Si and $^{26}$Al$^m$ with a very
low threshold and nearly 100\% overall efficiency, so superficially one might conclude that $\epsilon_1/\epsilon_2=1$ and
that the efficiency ratio can consequently be ignored.  However, the end-point energy of the $\beta^+$ spectrum
corresponding to the $^{26}$Si superallowed transition is 3819 keV, while that for the equivalent transition from
$^{26}$Al$^m$ is 3211 keV.  The 608-keV difference between them means that the shapes of the two spectra are slightly
different from one another.  Furthermore, the Gamow-Teller branches that occur only for the $^{26}$Si decay contribute
to the differences as well.  This is illustrated
by the calculated $\beta^+$ spectra shown in Fig.\,\ref{fig4}(a).

Because we operate our gas detector in the plateau region (see Sec.\,\ref{gasc}), we can be confident that our electronic
detection threshold is too low for losses from that source to have any effect on our overall detection efficiencies.  
However, the aluminized Mylar tape (half-thickness, 38 $\mu$m) and the Havar window of each detector gas cell (1.5 $\mu$m
thick) stop the lowest-energy positrons, thus preventing them from reaching the active volume of the detector.  In effect, 
this imposes a low-energy cut-off and, since the parent and daughter $\beta$ spectra have slightly different shapes at low
energies, the fraction of positrons lost in one case is slightly different from that in the other.  For a pair of decays
like ours, where the average positron energy is greater for the parent than it is for the daughter, the ratio is always
$\epsilon_1/\epsilon_2\geq1$. 

Since the decay positrons are emitted isotropically, their paths through the tape and window cover a range of lengths, 
resulting in a low-energy cut-off that is not sharp.  Nevertheless, for any given cut-off energy the effect on the efficiency
ratio can readily be calculated from the known $\beta$-spectrum shapes.  Using the Monte Carlo code EGSnrc (version V4-r2-3-0)
\cite{Ka00}, in which we modeled our exact tape/window/detector geometry, we obtained the overall system efficiency as a
function of positron energy, the result being shown in Fig.\,\ref{fig4}(b).  Note that the code was only required to calculate
the losses due to the tape and window.  We have extensively tested \cite{Go08} the accuracy of three Monte Carlo codes --
EGSnrc, Geant4 and PENELOPE -- in fitting experimental results from a thin scintillation detector for low energy conversion
electrons as well as $\beta$-decay spectra.  We found that EGSnrc offers the best combination: it agrees well with experiment
and it operates most efficiently.

The Monte Carlo result was then integrated with each of the two spectra in Fig.\,\ref{fig4}(a) to obtain our overall
efficiencies for detecting the parent and the daughter activities.  From these, we derived the efficiency ratio, 
$\epsilon_1/\epsilon_2 = 1.00143(25)$, where we have assigned an uncertainty that encompasses a $\pm10$\% relative
uncertainty in the calculated ranges.  This uncertainty was based on an assessment of how much the ratio would change
if the source were not located at exactly the center of the detector, thus slightly changing the losses in the tape and
window.  However, by eliminating all cycles in which the ratio of recorded positrons to implanted silicon ions was
anomalously low, we ensured that all analyzed cycles corresponded to central or nearly central source locations.  The
uncertainty we have assigned to the efficiency ratio is a very generous allowance for the range of locations allowed
in the accepted cycles.  The value we obtained for $\epsilon_1/\epsilon_2$ and its uncertainty were subsequently used
in the analysis of all runs. 

\begin{figure}
  \includegraphics[width=7.0cm]{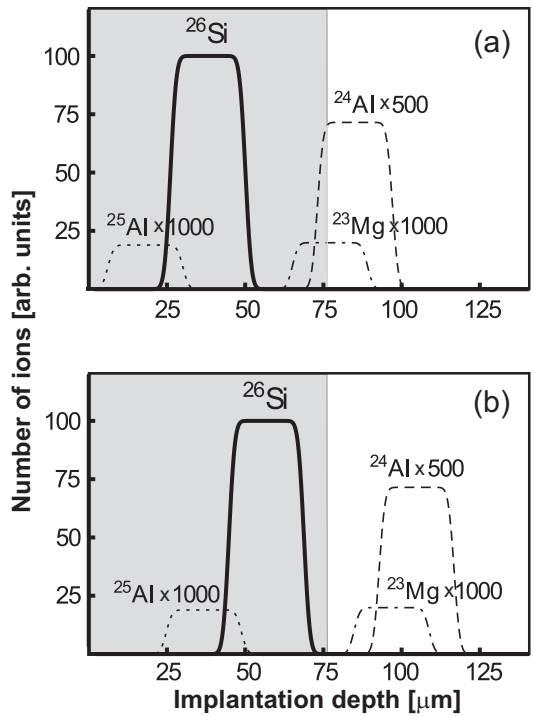}
 \caption{\label{fig5} Illustrations of calculated implantation profiles in Mylar for the $^{26}$Si
beam and those impurities with similar ranges.  All beams enter from the left.  The top illustration (a)
gives the profile after the beams have passed through 159 $\mu$m of aluminum, while the bottom one (b) gives
the result after 171 $\mu$m of aluminum.  The shaded region in both illustrations corresponds to the
actual thickness of our collection tape.  Those ions within the shaded region are collected in our sample;
all others are not.}
\end{figure} 

\subsection{Sample impurities \label{sub:imp}}

From the position-sensitive detectors that we inserted into the MARS focal plane before the measurement and periodically during 
it, we could identify and monitor any reaction products that might potentially contribute as impurities to our
selected $^{26}$Si beam.  Only three were detectable and all were extremely weak: $^{25}$Al (0.02\% of the
$^{26}$Si intensity), $^{24}$Al (0.14\%) and $^{23}$Mg (0.02\%).  However, even these small percentages must be
reduced when we consider what was actually retained in the aluminized Mylar tape that transported
the collected activity to our detector. As the impurity ions passed through our aluminum degraders, each impurity
experienced a different energy loss from the others and from $^{26}$Si.  The result is illustrated in
Fig.\,\ref{fig5} for the two thicknesses of degraders employed during the measurement (see Sec.\,\ref{exp}).

Figure \ref{fig5} presents the results of calculations based on the SRIM code \cite{Zi08}.  Before the main measurements
began, we recorded the collected $^{26}$Si activity as a function of aluminum degrader thickness and were thus
able to determine experimentally the precise thickness required to center the $^{26}$Si deposit in the tape.  The value
obtained was very close to that predicted by SRIM, which gives confidence that the calculations can be relied on to
determine the spatial distributions of the impurities relative to that of $^{26}$Si.  From Fig.\,\ref{fig5} it is
evident that only 10\% of the $^{24}$Al and 50\% of the $^{23}$Mg was collected in the tape when the $^{26}$Si was centered
in the tape (top illustration) and none at all when the $^{26}$Si was placed near the back of the tape (bottom).  
Consequently, relative to the collected $^{26}$Si activity, the collected $^{24}$Al and $^{23}$Mg activities were
both approximately 0.01\% when the $^{26}$Si was centered; they were zero when it was near the back.  The only remaining
impurity, $^{25}$Al, was fully collected at both degrader settings, but its relative intensity was only 0.02\% to
start with.  The effect of these three impurities was only barely significant, but we included them with their
appropriate relative intensities when we fitted the collected decay spectra.  We also incorporated a $\pm$30\% relative
uncertainty on the intensities used.

\subsection{Time decay analysis \label{sub:tdecay}}
 
We fitted the data from each of the 55 runs separately, incorporating four components: $^{26}$Si; its daughter,
$^{26}$Al$^m$; the weak impurities, $^{25}$Al, $^{24}$Al and $^{23}$Mg; and a constant background.  The half-life
of $^{26}$Al$^m$ was fixed at its known value of 6345.0(19) ms \cite{Ha09}.  Its initial activity relative to that
of $^{26}$Si was set for each run to the value obtained from numerical integration of the measured time-profile
for the collected $^{26}$Si beam in that run (see Sec.\,\ref{sub:pd} and Fig.\,\ref{fig3}); and the efficiency
ratio, $\epsilon_1/\epsilon_2$, was fixed at the value established in Sec.\,\ref{sub:eff}.  The half-lives of
$^{25}$Al, $^{24}$Al and $^{23}$Mg were taken to be 7.182(12)\,s \cite{Se08}, 2.053(4)\,s \cite{En90} and
11.324(10)\,s \cite{Se08} respectively.  Their initial relative activities were obtained from the measurement
and calculations described in Sec.\,\ref{sub:imp}.

\begin{figure}[t]
  \includegraphics[width=\columnwidth]{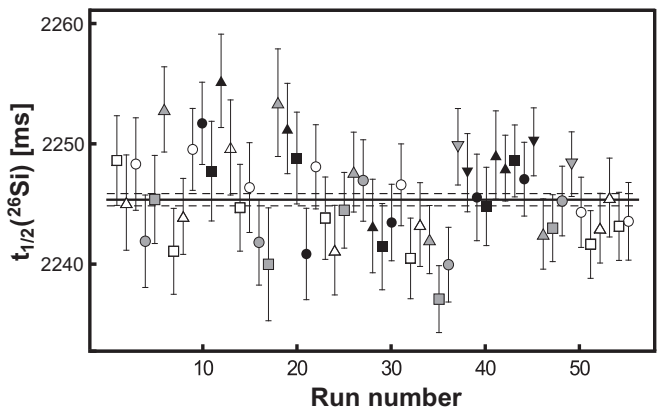}
 \caption{\label{fig6} Test for possible systematic bias in the $^{26}$Si half-life measurement due to
discriminator threshold or detector voltage.  Open/grey/black symbols represent the three discriminator settings, 
150\,mV/200\,mV/250\,mV; the four detector biases, 2550\,V, 2650\,V, 2750\,V and 2850\,V are represented by the
symbol shapes $\triangle$, $\square$, {\Large$\circ$} and $\triangledown$, respectively.  The average value for the
half-life is 2245.26(51) ms (statistical uncertainty only) with $\chi^2/ndf=70.6/54$. The average value appears
as the solid line, with dashed lines as uncertainty limits.}
\end{figure}

\begin{figure}[t]
  \includegraphics[width=\columnwidth]{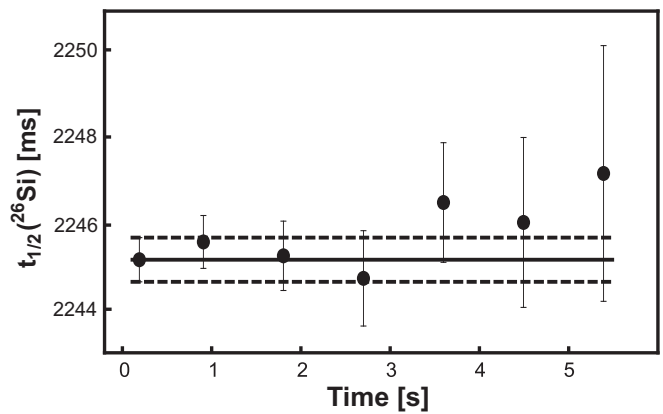}
 \caption{\label{fig7} Test for possible systematic bias in the $^{26}$Si half-life measurement caused by undetected
short-lived impurities or by rate-dependent counting losses.  Each point is the result of a separate fit to the data;
the abscissa for each point represents the time period at the beginning of the counting cycle for which the data were
omitted from that particular fit.  The solid and dashed lines correspond to the average half-life value and
uncertainties given in Fig.\,\ref{fig6}.}
\end{figure}

Since each run was obtained with a different combination of detection settings, we could use the individually fitted
$^{26}$Si half-lives to test for any systematic dependence on those settings. As seen in Fig.~\ref{fig6}, the half-life
results show no systematic dependence on detector bias or discriminator threshold. Although not illustrated, the
results were also found to be independent of both the imposed circuit dead time and the length of time for which the
sample was collected.  As a final systematic check, in this case for the possible presence of short-lived impurities or
other possible count-rate dependent effects, we removed data from the first 0.9$\,$s of the counting period in each
measurement and refitted the remainder; then we repeated the procedure, removing the first 1.8$\,$s, 2.7$\,$s and so
on.  From Fig. \ref{fig7} it can be seen that, within statistics, the derived half-life was also stable against these
changes.
 
With these possible systematic effects eliminated as significant factors, we can combine all 55 runs to obtain a value
for the $^{26}$Si half-life of $t_{1/2}$($^{26}$Si) = 2245.26\,ms, with a statistical uncertainty of $\pm$0.51\,ms.  The
normalized $\chi^2$ of this average is 1.31 and the statistical uncertainty has been multiplied by the square root of this
number.  The result itself represents a self-consistent analysis of about 200 million combined $^{26}$Si and $^{26}$Al$^m$
decay events.
 
This analysis has depended upon the source being centrally located in the gas counter for each cycle since serious
misplacement would have resulted in a different parent-daughter relative efficiency for that cycle (see Sec.\,\ref{sub:eff}).
We ensured centrality by rejecting cycles that had anomalously low ratios of recorded positrons to implanted silicon ions.
For each run we included only those cycles with ratios between 91 and 100\% of the maximum value obtained for that run.  
To test whether our result is in any way sensitive to this choice, we repeated the full analysis for subgroups of the
cycles corresponding to ratios between 98-100\%, 96-98\%, 94-96\% and so on.  Although having larger uncertainties, the half-lives
obtained from those subgroups within our selected band of 91-100\% were all statistically consistent with our quoted result of
2245.3(5)\,ms.  More importantly, the nearest group of rejected cycles -- those with ratios in the range 88-90\% --
were also consistent, yielding a value of 2246.7(30)\,ms.  Clearly, our selection criterion for accepting cycles was a
conservative one that does not adversely affect the result.

It is also interesting to compare this result with the half-life value derived from a fit to the data that does not
impose a fixed link between the parent and daughter activities.  We described in Secs.\,\ref{sub:pd} and \ref{sub:eff}
how we could reduce by one the number of free parameters in the fit by independently determining the ratios $N_2/N_1$ and
$\epsilon_1/\epsilon_2$.  This improved the statistical uncertainty in the fitted result but did introduce an additional
uncertainty associated with the ratio $\epsilon_1/\epsilon_2$ (see the second line in Table\,\ref{table1}).  If we ignore
the parent-daughter link and fit the data with four variable parameters (including background), we obtain a half-life result
of 2243.2(22) ms, which is statistically consistent with the value presented in Table\,\ref{table1} but has a thrice
larger uncertainty.

\subsection{Uncertainty budget \label{sub:unb}}

There are other contributions, of course, to our final uncertainty beyond the contribution from counting statistics.  
We itemize them all in Table \ref{table1}.  Counting statistics is the largest contributor to the overall
uncertainty, but the contribution associated with the different efficiencies for detecting parent and daughter
activities is a close second.  Less important contributors were the uncertainties in the $^{26}$Al$^m$ half-life,
the dominant circuit dead-time and the weak sample impurities.  Our final result for the $^{26}$Si half-life is
2245.3(7)\,ms, in which statistical and systematic uncertainties have been combined.

\begin{table}[b]
\caption{Error budget for the $^{26}$Si half-life measurement.
\label{table1}}
\vskip 1mm
\begin{ruledtabular}
\begin{tabular}{lc}
Source & Uncertainty (ms)  \\
\hline \\[-2.5mm]
statistics		&0.51	\\
efficiency ratio, $\epsilon_1/\epsilon_2$      &0.37  \\
$t_{1/2}(^{26}$Al$^m$)	&0.16	\\
dead time              &0.07   \\
sample impurities	&0.04	\\[+2mm]

Total	& 0.66    \\[+2mm]
$^{26}$Si half-life result (ms)  &2245.3(7)  
\end{tabular}
\end{ruledtabular}
\end{table}

\subsection{Comparison with previous results \label{sub:comp}}

Three previous measurements of the $^{26}$Si half-life are listed in the most recent review \cite{Ha09}: 2210(21) ms
\cite{Ha75}, 2240(10) ms \cite{Wi80} and 2228.3(27) ms \cite{Ma08}.  The first two of these results were recorded more
than 30 years ago and have uncertainties larger by more than a factor of 10 than our present result; one of them agrees
with our result, while the other lies one-and-a-half of its standard deviations away.  This can hardly be viewed as a
matter of concern.  However, the last of the previous measurements was published only two years ago, has a quoted
uncertainty that is only four times ours and differs from our result by more than six times that uncertainty.  This
discrepancy has to be addressed.

In their measurement, Matea {\it et al.}~\cite{Ma08} employed a purified $^{26}$Si beam from the JYFLTRAP Penning-trap
facility and implanted it in a movable 100-$\mu$m-thick Mylar tape located at the center of a hollow cylinder whose
2-mm-thick walls were made from plastic scintillator optically coupled to two photomultipliers.  After sample implantation,
the decay positrons were counted in the plastic scintillator, after which the tape removed the sample and the cycle
was repeated.  The experimental arrangement is pictured in Ref.\,\cite{Be08}; there it can be seen that the tape enters
and exits the scintillator cylinder via the same opening and passes over a roller, which is located within the cylinder.

The method Matea {\it et al.}~used to analyze their data was similar to ours in that they fixed the activity of
$^{26}$Al$^m$ based on its calculated production from the decay of an initially pure $^{26}$Si sample.  Unfortunately
they did not take account of any detection-efficiency difference between the parent and daughter activities \cite{Bl09}
(see Sec.\,\ref{sub:eff}).  Furthermore, their experimental conditions make this a much more serious problem for them
than it was for us.  First, both their thicker tape and the roller they used to allow that tape to move introduced a
more significant and complicated low-energy cut-off due to positrons ranging out in those materials.  Second, with a
plastic scintillator and photomultiplier tubes generating their signals, they certainly needed to set a non-negligible
low-energy electronic threshold. Third, their scintillator was, to some extent, sensitive to $\gamma$ rays, which would
slightly favor detection of $^{26}$Si, which produces $\beta$-delayed $\gamma$ rays, while $^{26}$Al$^m$ does not.

Although it is obviously impossible for us to model in detail someone else's experimental apparatus, we approximated
their arrangement in the EGSnrc Monte Carlo code \cite{Ka00}, the same code we used to help determine the efficiency
ratio for our measurement (see Sec.\,\ref{sub:eff}).  We found that the results were very sensitive to just those
experimental details that we could not define precisely.  Nevertheless, we determined that the value of
$\epsilon_1/\epsilon_2$ could easily reach 1.005 or possibly larger.  If this ratio had been applied to their data, it
would have increased their measured half-life by $\sim$0.4\%, bringing it to 2237 ms, approximately half way to our
result, and undoubtedly with a much larger uncertainty than the 2.7 ms originally claimed.  Clearly this is only an
indicative not a definitive calculation, since the exact correction and its application to the data requires
information only available to the original authors.  For now, though, we believe that it reveals a serious omission in
the original analysis by Matea {\it et al.}~\cite{Ma08}, and that their published result should simply be discounted.

\section{Conclusions}
%
We report here the first measurement of the half-life of the superallowed $\beta^+$ emitter $^{26}$Si obtained
with a precision of better than 0.1\%. Since $^{26}$Si and its daughter $^{26}$Al$^m$ have comparable half-lives, this
measurement required us to use the technique we developed for the measurement of the $^{34}$Ar half-life
\cite{Ia06}, in which we link the parent and daughter decays based on a precise knowledge of the rate of deposition
of the $^{26}$Si source.  We also had to make small but important corrections for detection-efficiency differences
between the parent and daughter $\beta^+$ decays.  Possible sources of error were carefully investigated and an
error budget established.

Our new precise result for the $^{26}$Si half-life, 2245.3(7) ms, is considerably different from the average value
quoted in the most recent survey of world data for superallowed $0^+$$\rightarrow$\,$0^+$ $\beta$-decay transitions
\cite{Ha09}.  However, the average there is dominated by a measurement \cite{Ma08} that we now argue is flawed and
should be discarded, since it leaves out the correction for parent-daughter detection-efficiency differences.  

With our half-life result for $^{26}$Si decay having 0.03\% precision, and the $Q_{EC}$ value for its superallowed
branch being recently determined to 0.0025\% \cite{Er09}, the $\F t$ value for the branch can in principle be
determined to sub-0.1\% precision if the branching ratio can be measured to that level.  This is a difficult but
potentially achievable goal and, if accomplished, would bring $^{26}$Si decay into the same category of precision as
the best known superallowed transitions.

This would allow a particularly interesting comparison with its mirror, $^{26}$Al$^m$ decay.  If calculated with
Saxon-Woods radial wave functions \cite{To08}, the nuclear-structure-dependent correction, ($\delta_C-\delta_{NS}$), 
for the $^{26}$Al$^m$ decay is 0.305(27)\% \cite{Ha09}, the smallest value for any superallowed transition, while
the equivalent correction term for the $^{26}$Si decay is 0.650(34)\%.  The difference between them, 0.345(43)\%, 
is reduced considerably, to 0.145(81)\%, if Hartree-Fock wave functions are used instead \cite{Ha09}.  An experimental
determination of the difference between these mirror transitions with sub-0.1\% precision would discriminate
between these theoretical approaches and could potentially reduce the theoretical uncertainties that affect the
ultimate determination of $V_{ud}$ and the unitarity of the CKM matrix.

\begin{acknowledgments}

This work was supported by the U.S. Department of Energy under Grant No.\,DE-FG03-93ER40773 and by the Robert
A. Welch Foundation under Grant no.\,A-1397. 

\end{acknowledgments}

\end{document}